# Cooperation After the Algorithm: Designing Human-AI Coexistence Beyond the Illusion of Collaboration

Tatia Codreanu[1*]

[1] Imperial College London, London SW7 2AZ, UK; I-X Imperial AI in Science Affiliate, London, UK; Information Cultures, Media Pedagogies, and Screen-Based Research Group, France

*Corresponding author: t.codreanu@imperial.ac.uk

**Abstract**

Generative artificial intelligence systems increasingly participate in research, law, education, media, and governance. Their fluent and adaptive outputs create an experience of collaboration. However, these systems do not bear responsibility, incur liability, or share stakes in downstream consequences. This structural asymmetry has already produced sanctions, professional errors, and governance failures in high-stakes contexts We argue that stable human-AI coexistence is an institutional achievement that depends on governance infrastructure capable of distributing residual risk. Drawing on institutional analysis and evolutionary cooperation theory, we introduce a formal inequality that specifies when reliance on AI yields positive expected cooperative value. The model makes explicit how governance conditions, system policy, and accountability regimes jointly determine whether cooperation is rational or structurally defective. From this formalization we derive a cooperation ecology framework with six design principles: reciprocity contracts, visible trust infrastructure, conditional cooperation modes, defection-mitigation mechanisms, narrative literacy against authority theatre, and an Earth-first sustainability constraint. We operationalize the framework through three policy artefacts: a Human-AI Cooperation Charter, a Defection Risk Register, and a Cooperation Readiness Audit.

Together, these elements shift the unit of analysis from the user-AI dyad to the institutional environment that shapes incentives, signals, accountability, and repair. The paper provides a theoretical foundation and practical toolkit for designing human-AI systems that can sustain accountable, trustworthy cooperation over time.

## 1. Introduction

GenAI now participates in professional and everyday environments with unprecedented speed. It writes, analyzes, summarizes, and advises. The experience of dialogue gives many users the impression of cooperation. Nevertheless, the structural conditions for cooperation (Ostrom, 1990; Ostrom et al., 1992) are not currently satisfied. LLMs optimize for predictive coherence. They do not optimize for mutual benefit, share responsibility for outcomes, or face consequences when errors occur. Instead, human users and institutions bear the full weight of downstream failures[1].

Recent work in AI alignment (Bai et al. 2022; Ouyang et al. 2022) and governance (Bommasani et al. 2021) has begun to address these challenges, but approaches remain primarily technical. Parallel efforts have outlined principles for designing AI systems that serve social good (Floridi et al., 2020), however these frameworks do not specify the institutional infrastructure required to sustain cooperative human-AI relationships over time. Empirical studies show that explanations and transparency alone do not solve the over-reliance problem when institutional pressures remain unchanged (Bansal et al. 2021). This paper moves beyond diagnosing the asymmetry to propose a design framework for creating genuine, accountable cooperation ecologies.

---

[1] American Bar Association. (2024). *Formal Opinion 512: Generative Artificial Intelligence Tools;* European Union. (2024). *Regulation (EU) 2024/1689 laying down harmonised rules on artificial intelligence (Artificial Intelligence Act);* Federal Court of Australia. (2025). *Artificial intelligence use in the Federal Court of Australia* (Notice to the Profession, 29 April 2025); Judicial Office (UK). (2023). *Artificial Intelligence (AI) Guidance for Judicial Office Holders; Mata v. Avianca, Inc.*, 678 F. Supp. 3d 443 (S.D.N.Y. 2023); Solicitors Regulation Authority. (2023). *Risk Outlook: The use of artificial intelligence in the legal market.*

Our central thesis is the following: stable human-AI cooperation is impossible without institutional infrastructure that distributes residual risk.

The prevailing user-AI dyad is insufficient; sustainable cooperation requires a triadic structure of *User-AI-Institution*, where institutions provide the rules, monitoring, and repair mechanisms that make accountability possible.

We provide a formal model of this claim and derive six actionable principles for designing such institutions.

## 2. The Illusion of Cooperation: A Formal Model

Humans evolved to read communicative signals as evidence of inner states and intentions, and to rely on such cues in social coordination (Darwin, 1872; Grice, 1975).

Social coordination depends on interpreting cues of confidence and intent (Merleau-Ponty, 1962). Generative systems operate with exceptional fluency in this signalling domain. Their responsiveness can trigger systematic over-attribution of competence and authority. Under conditions of time pressure and institutional constraint, the path of least resistance is to treat fluent output as reliable guidance.

These pressures produce what we term the *illusion of cooperation*: users experience responsiveness and apparent alignment, but the structural conditions for accountable cooperation are absent. When errors occur, the asymmetry becomes visible. The model does not face disciplinary review, bear liability, or incur sanctions. Human users and institutions absorb the consequences[2].

---

[2] Judicial Office (UK) (2023) *Artificial Intelligence (AI) Guidance for Judicial Office Holders*.

## 2.1 A Formal Inequality

For reliance on AI to be justified, and for institutions to reasonably expect its use, the expected net value of cooperation must be positive (Eq. 1):

(1) $E[\text{Net Cooperation Value}] = \text{VOI}(\omega, g, \pi_A) - C_{\text{interaction}}(u, d) - L_{\text{residual}}(a, \omega; \Lambda) > 0$

Where:
- $\text{VOI}(\omega, g, \pi_A)$ (Value of Information) represents the expected epistemic gain from consulting the system, given task $\omega$, governance conditions $g$, and AI policy $\pi_A$. This includes improved recommendations, calibrated uncertainty, and decision-relevant insight.
- $C_{\text{interaction}}(u, d)$ is the cost of consultation: time, cognitive effort, and workflow disruption for user $u$ in domain $d$.
- $L_{\text{residual}}(a, \omega; \Lambda)$ is a critical term. It captures the expected liability or downside that remains with human actors after the AI's contribution. The parameter $a$ represents the action or output of the AI. The parameter $\Lambda$ represents the *accountability regime*, the set of institutional rules that determine who bears responsibility for errors. In current deployments, $\Lambda$ assigns nearly all residual risk to the human and institution. The system contributes insight but does not share consequences. For many high-stakes applications, this inequality remains negative: expected gains are uncertain, interaction costs persist, and residual liability remains concentrated on human actors.

Institutions choose governance conditions $g$ and accountability regime $\Lambda$, system designers choose policy $\pi_A$, users select actions conditioned on $u$ and $d$ and tasks ω arrive from the environment.

In the case *Mata v. Avianca, Inc.*[3] the lawyer who experienced the interaction as cooperative, the LLM was fluent, responsive, and seemingly helpful. But because the accountability regime Λ places all the downside on the lawyer, the structural conditions for real cooperation were absent. The AI had no 'skin in the game'. It could defect (provide a hallucination) without facing any consequences, while the human cooperator (the lawyer) bears the full brunt of the defection.

---
[3] *Mata v. Avianca, Inc.,* 678 F. Supp. 3d 443 (S.D.N.Y. 2023)

The AI system, as a predictive cooperator, faces no consequences; it cannot be sanctioned by a bar association or held in contempt of court. Consequently, the residual liability term, $L_{\text{residual}}(a, \omega; \Lambda)$ becomes catastrophically large, encompassing professional sanctions, legal penalties, and irreparable reputational damage. Even though the interaction felt cooperative, the expected net value of cooperation is rendered overwhelmingly negative, exposing the hallucination as a structural failure of the accountability infrastructure that makes genuine cooperation possible.

This paper suggests that we need to change $\Lambda$. We need to build institutional infrastructure (mandatory verification logs, clear accountability leads, and audit trails). We cannot make the AI 'feel' responsibility, we cannot code intent, but we can change the rules of the game so that the organizations deploying the AI share more of the $L_{\text{residual}}(a, \omega; \Lambda)$ making the overall equation more balanced and the cooperation more sustainable.

### *2.2 From Diagnosis to Design*

This formulation clarifies two possibilities. First, when governance conditions are adequate ($g \geq g^*$) and AI performance is demonstrably calibrated, the inequality may become robustly positive across repeated uses under comparable conditions.

In the condition $g \geq g^*$, the symbol $g$ represents the prevailing governance conditions surrounding a human-AI deployment. The threshold $g^*$, signifies the minimum level of institutional and governance quality required to make the cooperation inequality potentially positive.

In other words, $g^*$ is the point on the governance spectrum where the surrounding infrastructure (rules, monitoring, accountability mechanisms) is adequate enough that the human actor's residual liability $L_\text{residual}(a, \omega; \Lambda)$ becomes manageable, and the value of information $\text{VOI}(\omega, g, \pi_A)$ from the AI can be trusted.

Higher $g$ increases VOI via better calibration and verification pathways and decreases $L_\text{residual}$ via monitoring, traceability, and faster repair.

A conditional cooperator will rationally use the system. When such conditions hold, institutions may legitimately encode an expectation of consultation. A *duty to consider* emerges as a structural consequence of an ecology in which consultation improves outcomes and governance distributes risk.

Designing human-AI coexistence therefore requires engineering the conditions under which this inequality holds for defined classes of decisions. Cooperation must be supported by calibrated signals, institutional accountability, and mechanisms for repair.

## 3. A Framework for Cooperation Ecology

Cooperation is a process sustained by repeated interaction, credible signalling, shared norms, and mechanisms for monitoring and repair (Alexander, 1987; Axelrod &Hamilton, 1981; Boyd & Richerson, 1985, 1989; Ostrom, 1990; Ostrom et al.1992). Ostrom's (1990) institutional analysis of commons governance shows that stable cooperation requires clear rules, transparency, graduated sanctions, and ways to correct errors. We adapt this framework to human-AI systems, treating interaction as an *ecology* composed of incentives (Trivers, 1971), signals (Ostrom, 1990), decision conditions, and formal and informal institutions (Ostrom, 1990).

### *3.1 Taxonomy of Cooperation Roles*

Building on cooperation theory and institutional analysis (Axelrod and Hamilton 1981; Ostrom 1990; Boyd and Richerson 1985), this section introduces a taxonomy of roles that shape how cooperation emerges in human-AI systems. Each role corresponds to a distinct parameter movement within the cooperation inequality and will be defined in relation to governance conditions, information value, and residual risk.

**Responsible cooperators**: Humans who verify outputs, remain accountable for decisions, and maintain repair mechanisms (actively verifying claims and maintaining escalation pathways). Responsible cooperators increase the effective governance level *g* within their local context and reduce $L_{\text{residual}}(a, \omega; \Lambda)$ through earlier detection and mitigation of error.

**Predictive cooperators**: AI systems that generate contextually useful outputs aligned with user intent but without bearing responsibility. Their contribution operates primarily through $\text{VOI}(\omega, g, \pi_A)$. Improvements in calibration, uncertainty signalling, and refusal boundaries under policy $\pi_A$ increase the expected value of information without altering the allocation of liability.

**Over-attributors**: Users who, influenced by AI fluency and institutional pressure, treat predictive outputs as authoritative beyond their statistical reliability. Over-attribution reduces effective verification and increases the probability of uncorrected error, thereby increasing $L_{\text{residual}}(a, \omega; \Lambda)$ and rendering the cooperation inequality more likely to become negative even when nominal governance structures exist.

**Structural defectors**: Organizations that deploy systems without adequate governance infrastructure, such as audit trails or accountability pathways, shifting risk to end-users while preserving corporate performance incentives. Structural defection keeps governance conditions below the threshold $g^*$ and maintains an accountability regime $\Lambda$ that concentrates residual liability on individual users. This practice constitutes *responsibility laundering*, the systematic transfer of accountability from organizations to individuals.

Cooperation breakdown is a predictable consequence of parameter misalignment. When structural defectors suppress $g$ and over-attributors amplify $L_{\text{residual}}$, the expected net value of cooperation turns negative. When responsible cooperators and calibrated predictive cooperators operate within governance conditions at or above $g^*$, the inequality can become robustly positive.

The category of structural defectors is crucial for policy. These actors face strong market incentives to externalize risk: speed and profitability favour minimal oversight, while liability remains with users. Countering structural defection requires regulatory mechanisms that make the full costs of deployment visible and assign them to decision-makers.

## 4. Results: Six Principles for Cooperative Alignment

We introduce *Cooperative Alignment*, a framework integrating principles from evolutionary cooperation theory, Ostromian commons governance, and sustainability science. Each principle combines a theoretical claim with an implementable design requirement.
The six principles that follow can be understood as institutional interventions designed to modify the parameters $g$, $\pi_A$ and $\Lambda$ so that the cooperation inequality becomes positive under defined task conditions $\omega$.

### *4.1 Principle 1: Reciprocity Without Symmetry*

**Theoretical claim:** Cooperation does not require equal capability, but it does require reliable mutual benefit over time (Alexander, 1987; Axelrod & Hamilton, 1981). This *functional reciprocity* must be engineered explicitly.
**Design requirements:** Define a *reciprocity contract* specifying:
- What the system owes the user (e.g., uncertainty signalling, refusal boundaries, traceability)

- What the user or deploying organization owes the system's ecosystem (e.g., oversight, lawful purpose, safe workflows)

**Hypothesis:** Contexts with explicit reciprocity contracts will show higher rates of appropriate reliance and lower rates of automation bias compared to contexts without such contracts.

### 4.2 Principle 2: Institutions Make Trust Real

**Theoretical claim:** Trust emerges from rules, monitoring, and repair. It is an institutional outcome (Ostrom, 1990; Ostrom et al.1992, Sugden,1986).
**Design requirements:** Build visible *trust infrastructure* in high-stakes contexts:
- Immutable traceability logs (timestamped, non-editable records of prompts and outputs)
- Clear escalation paths for disputes or errors
- Incident response protocols
- Graduated sanctions for repeated misuse or negligent misconfiguration

**Hypothesis:** Deployments with visible trust infrastructure will generate higher user confidence and lower organizational risk exposure than those without.

### 4.3 Principle 3: Conditional Cooperation as the Default

**Theoretical claim:** The stable form of cooperation in human societies is conditional: individuals cooperate when others do, and defect when cooperation is not reciprocated (Alexander, 1987; Ostrom,1990).
**Design requirements:** System design must support *help*, *refusal*, and *non-assistance* modes (situations where assistance is withheld as inappropriate or harmful), dynamically triggered by context assessment of authority, consent, data quality, and oversight presence. This builds on technical work in constitutional AI and value alignment (Bai et al. 2022).
**Hypothesis:** Users interacting with systems that appropriately refuse harmful requests will develop more calibrated trust than users interacting with systems that always comply.

## 4.4 Principle 4: Defection is Ecological Damage to the Cooperation Environment

**Theoretical claim:** Defection degrades the shared environment for future cooperation (Ostrom, 1990; Ostrom et al.1992). In human-AI systems, defection includes both user misuse and organizational irresponsibility.

**Design requirements:** Define predictable *defection modes* (e.g., speed-over-safety incentives, responsibility laundering) and design countermeasures:

- Friction for high-risk actions (e.g., mandatory delays, additional verification steps)
- Mandatory human gates for irreversible decisions
- Transparency requirements that make defection visible to regulators and affected parties

**Hypothesis:** Systems with friction for high-risk actions will have lower rates of catastrophic errors than systems optimized for speed alone.

## 4.5 Principle 5: Narrative and Meaning are Cooperation Technology

**Theoretical claim:** Humans cooperate through shared stories and legitimacy signals, a phenomenon deeply connected to the mimetic dynamics that shape social learning and conflict (Girard, 1972). Girard's mimetic theory posits that human desire is not autonomous but is acquired by imitating the desires of others. In digital contexts, AI systems can become powerful models of desire and authority, leading users to mimic apparent confidence and trust in outputs. This is is due to the fact that the system performs authority so fluently. The mimetic dynamic transforms generated text into a target of uncritical imitation, enabling AI to generate what we term *authority theatre*: the performative display of expertise or control without substantive backing.

**Design requirements:** Build narrative literacy into user education and governance:

- Train users to recognize authority theatre

- Mandate honest communication of system limits (e.g., confidence scores, uncertainty estimates)
- Create clear contestability pathways for challenging outputs

**Hypothesis:** Users trained to recognize authority theatre will verify AI outputs more frequently and appropriately than untrained users.

### *4.6 Principle 6: Earth-First as the Top Constraint*

**Theoretical claim:** Cooperation depends on a stable ecological base. Computing infrastructure requires substantial energy and generates measurable carbon emissions (Crawford, 2021; Patterson et al., 2021)

If deployment undermines the ecological systems that sustain human societies, no cooperation framework can be durable. This is a classic *tragedy of the commons* problem: AI's environmental externalities are distributed across populations and time periods far removed from sites of profit extraction (Hickel et al. 2022; Rockström et al. 2009).

**Design requirements:** Adopt an Earth-first framing that subordinates AI deployment decisions to planetary boundaries:

- Implement energy and resource accounting
- Include environmental externalities in ROI calculations
- Prioritize deployments that reduce waste or harm
- Reject scaling of coercive or disinformative applications regardless of profitability

This principle operationalizes Ostrom's design principles for common-pool resources: clear boundaries on acceptable environmental impact, proportional equivalence between benefits and costs, and collective-choice arrangements that include affected communities.

**Hypothesis:** Organizations that publicly commit to Earth-first constraints will innovate more efficient, sustainable AI applications than those that do not.

## 5. Implications for Policy and Practice

In research, education, law, healthcare, and media, the key risk is the misreading of signals and the shifting of accountability. Policies should therefore: (a) treat AI outputs as non-authoritative drafts unless verified, (b) mandate traceability in high-stakes settings, and (c) explicitly assign responsibility for review and repair. Organizations should operationalize this framework through three artefacts:

1. **Human-AI Cooperation Charter:** Defines roles, conditions for use, and accountability lines. Includes a reciprocity contract specifying system and user obligations.
2. **Defection Risk Register:** Catalogues predictable failure modes (e.g., automation bias, responsibility laundering, prompt injection) and assigns mitigation owners with clear accountability.
3. **Cooperation Readiness Audit:** Evaluates whether a system's governance infrastructure justifies its deployment in a specific context. Includes threshold questions about reciprocity contracts, repair pathways, and environmental impact.

These artefacts translate theoretical principles into organizational practice. They create the institutional infrastructure that makes stable cooperation possible. These artefacts build on established approaches to algorithmic accountability. Raji et al. (2020) demonstrated that closing the AI accountability gap requires internal audit practices that go beyond one-off evaluations, a principle embedded in our Cooperation Readiness Audit's requirement for designated accountability leads with halt authority. Similarly, Metcalf et al. (2021) showed that effective algorithmic impact assessments must be embedded in institutional decision-making rather than treated as compliance checkboxes; our Defection Risk Register operationalizes this insight by assigning clear mitigation owners and requiring ongoing review of failure modes.

# 6. Discussion

## 6.1 Contributions

This paper makes three primary contributions. First, it provides a formal model of the cooperation asymmetry, making explicit the conditions under which human-AI interaction yields net positive value. Second, it introduces a cooperation ecology framework that shifts analysis from the user-AI dyad to the institutional infrastructure shaping incentives and accountability. In the dyad, $g$ is low and $\Lambda$ is concentrated on the user. In the triad, institutions set $g \geq g^*$ and distribute $\Lambda$ across accountable roles.
Third, it offers six design principles and three policy artefacts that operationalize the framework for practitioners and regulators.

## 6.2 Limitations

This framework requires empirical validation through controlled deployment studies. The principles are design heuristics; their operationalization will vary across contexts. The framework assumes rational institutional actors capable of implementing cooperation infrastructures, but many deployment contexts lack this capacity. Power asymmetries between AI developers and affected communities are not fully addressed. Cost implications for monitoring infrastructure may create barriers for smaller organizations.

## 6.3 Future Work

Future research will include: (a) empirical case studies testing specific principles (e.g., the effect of refusal modes on user trust), (b) refinement of the formal model to incorporate dynamic learning and adaptation, (c) exploration of enforcement mechanisms to ensure compliance with cooperation principles, and (d) investigation of global equity in access to cooperation infrastructure.

## *7. Conclusion*

AI systems currently create the *experience* of cooperation without the *architecture* for sharing responsibility. Stable, sustainable coexistence requires deliberate alignment of incentives, signals, decision environments, and institutional capacity for repair. Cooperation becomes rational once the expected cooperative surplus exceeds the expected cost of strategic and epistemic risk. When cooperation functions as a designed ecology embedded within institutional infrastructure, human-AI systems can develop trustworthiness, contestability, and long-term sustainability.

**Author Contributions**

The author conceived the framework, developed the taxonomy, and wrote the manuscript.

**Competing Interest Statement**

The author declares no competing interest.

**Appendix: Policy Artefact Templates**

*A1. Human-AI Cooperation Charter (Template Excerpt)*

| Section | Content |
|---|---|
| **System Identity & Version** | [Model/System Name, Version] |
| **Intended Domain of Use** | [e.g., Preliminary Legal Research, Academic Ideation] |
| **Reciprocity Contract** | System Obligations: Signal uncertainty when confidence < [threshold]; refuse tasks involving [list]; provide provenance trace for all outputs. User Obligations: Verify all factual claims; not use output for final decisions without expert review; report errors via [channel]. |
| **Accountability Lead** | [Role/Title] responsible for incident response and charter updates. |

*A2. Defection Risk Register (Template Excerpt)*

| Risk Mode | Description | Likelihood | Impact | Mitigation Owner |
|---|---|---|---|---|
| Responsibility laundering | Organization uses AI to obscure human accountability for harmful outcomes. | Medium | High | Legal/ Compliance Dept |

| Over-Attribution Cascade | Team members mimic each other's uncritical acceptance of AI output. | High | Medium | Team Manager; requires audit trail review |
| Automation Bias | User accepts incorrect AI recommendation without verification. | High | High | Training Lead; implements mandatory verification steps |

### *A3. Cooperation Readiness Audit (Checklist Excerpt)*

- Has a reciprocity contract been codified for this use case?
- Is there a working, low-friction error reporting and repair pathway?
- Does the deployment include visible trust infrastructure (traceability logs, escalation paths)?
- Have users been trained to recognize system limits and failure modes?
- Does the deployment reduce net resource use or environmental harm compared to alternatives?
- Is there a designated accountability lead with authority to halt deployment if risks emerge?